\documentclass{PoS}

\title{Yang-Mills theory on noncommutative space: does it exist?}

\ShortTitle{Does NCYM exist?}

\author{\speaker{Masanori Hanada}\\
 Yukawa Institute for Theoretical Physics, Kyoto University,\\
Kitashirakawa Oiwakecho, Sakyo-ku, Kyoto 606-8502, Japan\\
Stanford Institute for Theoretical Physics,
Stanford University, Stanford, CA 94305, USA\\
The Hakubi Center for Advanced Research, Kyoto University,\\
Yoshida Ushinomiyacho, Sakyo-ku, Kyoto 606-8501, Japan\\
        E-mail: \email{hanada@yukawa.kyoto-u.ac.jp}}

\abstract{
I revisit a basic question about the noncommutative Yang-Mills theory: if it exists or not,  
or more precisely, whether a nonperturbative formulation exists. 
As the most promising approach, I consider a formulation based on matrix models. 
It is explained that the existence of the noncommutative Yang-Mills theory is closely related to the Eguchi-Kawai equivalence. 
I argue that supersymmetric noncommutative Yang-Mills theory can be defined straightforwardly. 
Non-supersymmetric theories, such as QCD and pure bosonic theories, 
can presumably be defined, by modifying the ultraviolet and infrared behaviors appropriately. 

}

\FullConference{Proceedings of the Corfu Summer Institute 2015 "School and Workshops on Elementary Particle Physics and Gravity"\\
		1-27 September 2015\\
		Corfu, Greece}

\begin{document}
\section{Introduction}

The existence of Yang-Mills theory on noncommutative space is a highly nontrivial issue. 
Even the perturbative renormalizability is controversial due to the UV/IR mixing in the non-planar sector \cite{Minwalla:1999px}. 
The UV/IR mixing problem does not exist for certain cases, 
for example the planar limit \cite{GonzalezArroyo:1982ub,GonzalezArroyo:1982hz}
and maximally supersymmetric gauge theories \cite{Matusis:2000jf,Hanada:2014ima}. 
In such cases the noncommutative theory can be renormalized just in the same way as in the commutative theory. 
In this talk I consider more generic theories, 
based on a nonperturbative formulation motivated by string theory. 
I do not intend to discuss the renoemalizability; I will consider the very first step of the formulation: 
the stability of the noncommutative space. 

In type II string theory, D-branes couple to Ramond-Ramond (RR) fields and can be polarized nontrivially. 
With a certain background RR-flux, D$p$-branes can form a fuzzy-spherical D$(p+2)$-brane. 
In terms of $(p+1)$-dimensional supersymmetric U$(N)$ Yang-Mills theory which describes the low-energy dynamics of $N$ D$p$-branes, 
the action is modified due to the RR-flux so that fuzzy sphere configurations become classical solutions. 
This is the Myers effect \cite{Myers:1999ps}. 
By taking the large-$N$ limit while tuning the flux parameter with $N$, 
the fluctuations about this solution describe noncommutative Yang-Mills theory 
on the Moyal plane \cite{Aoki:1999vr}. 
Formally, the same mechanism can work in various large-$N$ gauge theories and matrix models.  
It would be reasonable to adopt it as a definition of the noncommutative theory, 
because the original, commutative theory can be obtained when the commutative limit is continuous. 

In order for this approach to work, the noncommutative background (e.g. fuzzy sphere, fuzzy torus) must be stable at large-$N$. 
This is again a nontrivial issue. As I will explain in  Sec.~\ref{sec:TEK}, 
the stability of the noncommutative background is closely related to the Eguchi-Kawai equivalence \cite{Eguchi:1982nm}. 
Roughly speaking, the noncommutative background is stable when the Eguchi-Kawai equivalence holds. 
In fact, historically the first appearance of noncmmutative Yang-Mills theory was 
in this context \cite{GonzalezArroyo:1982ub,GonzalezArroyo:1982hz}, 
although the geometric interpretation as noncommutative space was not realized at that time. 
I will explain the connection between noncommutative Yang-Mills theory and the Eguchi-Kawai model in 
Sec.~\ref{sec:TEK} and Sec.~\ref{sec:collapse}.  
There, it is also explained how the noncommutative background can turn unstable. 
In Sec.~\ref{sec:SYM} and Sec.~\ref{sec:multi-adj}, I explain how the instabilities can be cured. 
\section*{Note Added in July 2016}
I have read M.~Van Raamsdonk's paper \cite{VanRaamsdonk:2001jd} again, for the first time in these several years, 
and realized that all the essence 
about the center symmetry breaking/restoration in the twisted Eguchi-Kawai model with or without massive adjoint fermions 
has been explained there already in 2001.  
It might not be so clear when you read Ref.~\cite{VanRaamsdonk:2001jd} for the first time, but if you read it again after reading this article and/or 
articles on heavy adjoint fermions such as \cite{Azeyanagi:2010ne}, what he meant must be apparent. 
\section{Eguchi-Kawai model}\label{sec:Eguchi_Kawai}
Wilson's lattice gauge theory\footnote{
Below the volume is assumed to be $\infty$. 
} in $d$ dimensions with U$(N)$ gauge group is given by the following action,
\begin{eqnarray}
S_{\rm W}
=
-\beta N
\sum_{\mu<\nu}
\sum_{\vec{x}}
{\rm Tr}\left(
U_{\mu,\vec{x}}
U_{\nu,\vec{x}+\hat{\mu}}
U_{\mu,\vec{x}+\hat{\nu}}^\dagger
U_{\nu,\vec{x}}^\dagger
\right),  
\end{eqnarray}
where $\mu,\nu=1,2,\cdots,d$, 
$U_{\mu,\vec{x}}$ is a unitary link variable which is related to the gauge field by $U_{\mu,\vec{x}}=e^{iaA_\mu(\vec{x})}$ ($a$: lattice spacing), 
and $\beta$ is the inverse 't Hooft coupling at the cutoff scale. 
The original version of the Eguchi-Kawai model \cite{Eguchi:1982nm} is obtained by reducing it to a single-site lattice, 
\begin{eqnarray}
S_{\rm EK}
=
-\beta N
\sum_{\mu<\nu}
{\rm Tr}\left(
U_{\mu}U_{\nu}
U_{\mu}^\dagger
U_{\nu}^\dagger
\right).  
\end{eqnarray}
There are only $d$ link variables in the Eguchi-Kawai model.

The Eguchi-Kawai equivalence states that these two theories are equivalent in the 't Hooft large-$N$ limit ($N\to\infty$ with $\beta$ fixed), 
{\it if the U$(1)^d$ symmetry in the Eguchi-Kawai model $U_\mu\to e^{i\theta_\mu}U_\mu$ is not spontaneously broken.} 
The `equivalence' means that translationally invariant  observables take the same expectation values up to $1/N$ corrections. 
For example, the Wilson loop $W=\frac{1}{N}{\rm Tr}\left(U_{\mu,\vec{x}}U_{\nu,\vec{x}+\hat{\mu}}U_{\rho,\vec{x}+\hat{\mu}+\hat{\nu}}\cdots\right)$
and its counterpart in the Eguchi-Kawai model $\tilde{W}=\frac{1}{N}{\rm Tr}\left(U_{\mu}U_{\nu}U_{\rho}\cdots\right)$ agree at large-$N$, 
\begin{eqnarray}
\langle W\rangle_{\rm W}
=
\langle\tilde{W}\rangle_{\rm EK},  
\end{eqnarray}
where $\langle\ \cdot\ \rangle_{\rm W}$ and $\langle\ \cdot\ \rangle_{\rm EK}$ stand for the expectation values 
evaluated in the Wilson's lattice gauge theory 
and the Eguchi-Kawai model, respectively. 
That the unbroken U$(1)^d$ symmetry is necessary can be understood as follows.  
In the lattice gauge theory, only loops (closed lines) can have nonzero expectation values, because of the gauge invariance. 
Open lines such as $W=\frac{1}{N}{\rm Tr}\left(U_{1,\vec{x}}U_{2,\vec{x}+\hat{1}}\right)$ must vanish. 
In the Eguchi-Kawai model, on the other hand, any line is a loop, because there is only one point. 
However the U$(1)^d$ symmetry can be used to distinguish the counterparts of loops and open lines, 
because $\tilde{W}$ transforms as $\tilde{W}\to e^{i\theta}\tilde{W}$, where 
$\theta=\sum_{\mu=1}^d\theta_\mu\times (\#\ {\rm of}\ U_\mu - \#\ {\rm of}\ U_\mu^\dagger)$. 
Hence the counterparts of open lines automatically vanish provided that 
the U$(1)^d$ symmetry is not broken. 

In fact the U$(1)^d$ symmetry is broken spontaneously for $d>2$ \cite{Bhanot:1982sh}. 
It can be seen by writing $U_\mu$ as $U_\mu=V_\mu D_\mu V^\dagger_\mu$,    
where $D_\mu={\rm diag}(e^{i\theta_\mu^1},e^{i\theta_\mu^2},\cdots,e^{i\theta_\mu^N})$ and $V_\mu=e^{ia_\mu}$, 
and integrating out the `fluctuation' $a_\mu$ at one-loop level. Then the one-loop effective action, 
which is justified when the phases $\theta_\mu^i$ are not localized too much, is  
\begin{eqnarray}
S_{1-{\rm loop}}(\theta_\mu^1,\cdots,\theta_\mu^N) 
=
(d-2)\sum_{i<j}\log
\left(
\sum_\mu\sin^2\left(
\frac{\theta_\mu^i-\theta_\mu^j}{2}
\right)
\right). 
\label{eq:bosonic_one_loop}
\end{eqnarray}
This shows that the phases $\theta_\mu^i$ tend to localize in a small region.  
Because the U$(1)^d$ symmetry is the shift symmetry of the distribution of $\theta_\mu^i$ modulo $2\pi$, 
when $\theta_\mu^i$ localize the U$(1)^d$ symmetry is broken. 

In analogy to string theory, $\vec{\theta}^i$'s describe D-branes distributed on the torus, 
which interact with each other  
via strings stretched between them. The one-loop potential (\ref{eq:bosonic_one_loop}) 
shows that strings pull D-branes to come close to each other, 
so that a single bunch of D-branes is formed. 

As we will see below, this instability of the U$(1)$-symmetric background is essentially the same as the instability of noncommutative backgrounds, which are used to formulate noncommutative Yang-Mills theory.

\section{Twisted Eguchi-Kawai model and noncommutative Yang-Mills}\label{sec:TEK}
In order to save the U$(1)^d$ symmetry, the `twisted' Eguchi-Kawai model 
\cite{GonzalezArroyo:1982ub,GonzalezArroyo:1982hz} has been proposed.\footnote{
The `quenched' Eguchi-Kawai model has also been proposed \cite{Bhanot:1982sh,Gross:1982at,Parisi:1982gp}. 
However for purely bosonic theory the quenching cannot save the U$(1)^d$ symmetry, as found in \cite{Bringoltz:2008av}. 
} 
As we will see, the twisted Eguchi-Kawai model provides us with a nonperturbative formulation 
of noncommutaive Yang-Mills theory. 
Its action is given by 
\begin{eqnarray}
S_{\rm TEK}
=
-\beta N
\sum_{\mu<\nu}
Z_{\mu\nu}
{\rm Tr}\left(
U_{\mu}U_{\nu}
U_{\mu}^\dagger
U_{\nu}^\dagger
\right),  
\end{eqnarray}
where $Z_{\mu\nu}=e^{2\pi in_{\mu\nu}/N}$, $n_{\mu\nu}=-n_{\nu\mu}\in{\mathbb Z}$ is the twist factor, 
which is essentially the same as the flux term introduced by Myers in the context of string theory.
Below I consider $d=4$ and 
\begin{eqnarray}
n_{\mu\nu}
=
\left(
\begin{array}{cc|cc}
0 & kn & 0 & 0\\
-kn & 0 & 0 & 0 \\
\hline
0 & 0 & 0 & kn\\
0 & 0 & -kn & 0
\end{array}
\right), 
\qquad
N=kn^2,   
\end{eqnarray}
where integers $k$ and $n$ are related to the gauge group of the noncommutative theory U$(k)$  
and cutoff scales.
The generalizations to more generic cases are straightforward. 

The classical solution can be obtained by using the shift matrix $S_n$ and the clock matrix $C_n$, 
which are $n\times n$ matrices given by 
\begin{eqnarray}
S_n
=
\left(
\begin{array}{ccccc}
0 & 1 & 0 & \cdots & 0\\
0 & 0 & 1 & \cdots & 0\\
\vdots & \vdots & & \ddots&\vdots\\ 
0 & 0 & 0 & \cdots & 1\\
1 & 0 & 0 & \cdots & 0
\end{array}
\right), 
\qquad
C_n
=
{\rm diag}(1,\omega,\omega^2\cdots,\omega^{n-1}), 
\qquad
\omega=e^{2\pi i/n}. 
\end{eqnarray}
By noticing 
$\omega C_n S_n = S_n C_n$, it is easy to confirm that 
\begin{eqnarray}
U^{\rm (0)}_1=\textbf{1}_k\otimes C_n\otimes\textbf{1}_n, 
\qquad
U^{\rm (0)}_2=\textbf{1}_k\otimes S_n\otimes\textbf{1}_n, 
\nonumber\\
U^{\rm (0)}_3=\textbf{1}_k\otimes \textbf{1}_n\otimes C_n, 
\qquad
U^{\rm (0)}_4=\textbf{1}_k\otimes \textbf{1}_n\otimes S_n 
\end{eqnarray}
satisfy 
\begin{eqnarray}
Z_{\mu\nu}
U_{\mu}^{\rm (0)}U_{\nu}^{\rm (0)}
=
U_{\nu}^{\rm (0)}U_{\mu}^{\rm (0)}, 
\end{eqnarray}
and hence minimize the action. In terms of the noncommutative space, this is the $k$-coincident fuzzy torus solution. 
Although this solution is not invariant under the full U$(1)^4$ transformation, it is still invariant under ${\mathbb Z}_n^4$ 
up to the U$(N)$ transformation. 

By writing as $U^{\rm (0)}_\mu=e^{ia\hat{p}_\mu}$ and zooming up a tangent space of the fuzzy torus $a\hat{p}_\mu\sim 0$
(more precisely, by considering the excitations localized at $a\hat{p}_\mu\sim 0$),  
the noncommutative plane $[\hat{p}_\mu,\hat{p}_\nu]=i\cdot\frac{2\pi n_{\mu\nu}}{Na^2}\equiv i\cdot (\theta^{-1})_{\mu\nu}$ is obtained. 
By using Hermitian matrices $A_\mu$, which is related to the unitary variable by $U_\mu=e^{iaA_\mu}$, the action becomes\footnote{
This is essentially the bosonic part of the IKKT matrix model \cite{Ishibashi:1996xs}, which will be considered later. 
}
\begin{eqnarray}
S_{\rm TEK}
=
-\frac{a^4\beta N}{4}{\rm Tr}\left(
[A_\mu,A_\nu]
-
i\cdot (\theta^{-1})_{\mu\nu}\cdot\textbf{1}_N
\right)^2
\label{eq:twisted_IKKT}
\end{eqnarray}
up to terms higher order in $a$. 
This matrix model, expanded about $A_\mu=\hat{p}_\mu$, 
reproduces the U$(k)$ noncommutative Yang-Mills theory to all order in perturbation, 
with the following identification \cite{Aoki:1999vr}:


\begin{eqnarray}
\begin{array}{c|c}
{\rm Matrix\ Model} & {\rm NCYM} \\
\hline\\
\frac{kna^2}{2\pi}
& \theta\\
{\rm matrix}\ \hat{x}_\mu\equiv\theta_{\mu\nu}\hat{p}_\nu 
& {\rm coordinate}\  x_\mu\\
{\rm matrix\ product}\ \hat{x}_\mu\hat{x}_\nu 
& {\rm Moyal\ product}\ x_\mu\ast x_\nu\\
i[\hat{p}_\mu,\ \cdot\ ]
& {\rm ordinary\ derivative}\ \partial_\mu\\
i[A_\mu,\ \cdot\ ]
& {\rm covariant\ derivative}\ D_\mu\\
\hat{a}_\mu\equiv A_\mu-\hat{p}_\mu 
& {\rm U}(k)\ {\rm gauge\ field}\ A_\mu(x)\\
a^4{\rm Tr}
& \int d^4x\ {\rm Tr}\\
\sqrt{n}\sim\frac{1}{a}
&  {\rm UV\ cutoff}\ \Lambda_{\rm UV}\\
\frac{1}{\sqrt{n}}\sim a
&  {\rm IR\ cutoff}\ \Lambda_{\rm IR}
\end{array}
\end{eqnarray}

In the twisted Eguchi-Kawai reduction \cite{GonzalezArroyo:1982ub,GonzalezArroyo:1982hz}, 
the 't Hooft large-$N$ limit ($\beta$ is fixed and hence $a$ is also fixed) with fixed $k$ is taken.  
Then the noncommutativity parameter $\theta=\frac{kna^2}{2\pi}$ goes to infinity, 
and hence only the planar diagrams survive, which is equivalent to the planar sector of the commutative theory. 

In this talk, I consider the limit of noncommutative Yang-Mills theory, 
in which the noncommutativity parameter  $\theta$ is fixed. 
Therefore $n=\sqrt{N/k}$ is tuned as $n\sim 1/a^2$. 

\section{Twist is not enough}\label{sec:collapse}
It turned out that, for sufficiently large $N$, the ${\mathbb Z}_n^4$ symmetry of  the twisted Eguchi-Kawai model is broken 
at intermediate coupling region $\beta_{c,H}< \beta < \beta_{c,L}$ \cite{Ishikawa_talk,Teper:2006sp,Azeyanagi:2007su}.
(Essentially the same phenomenon in a related theory was reported in \cite{Bietenholz:2006cz}.) 
Furthermore, the critical value at weak coupling side increases with $N$ as $\beta_{c,L}\sim N$ for fixed $k$ \cite{Azeyanagi:2007su}. \footnote{
It has been pointed out that this instability cab be avoided by tuning $k$  with $N$ \cite{GonzalezArroyo:2010ss}. 
Although this modification can save the twisted Eguchi-Kawai equivalence, 
it does not help noncommutative Yang-Mills because it changes the gauge group U$(k)$. 
}
This scaling of $\beta_{c,L}$ can be understood rather easily in the following manner. 
If one interprets the action (\ref{eq:twisted_IKKT}) as an effective action of D-branes and open strings, 
the ${\mathbb Z}_n^4$-broken vacuum corresponds to a bunch of D-branes localized to a point. 
There the open strings are short and can easily be excited.  
On the other hand, in the ${\mathbb Z}_n^4$-unbroken vacuum, D-branes are spread out and hence 
most open strings are long, heavy and cannot be excited much. 
Therefore, the former has more dynamical degrees of freedom and hence it is entropically favorable; 
just by counting the number of off-diagonal elements, 
we can easily estimate that the difference of the entropies is of order $N^2$, 
which can be seen as a quantum correction in the effective action of D-branes by integrating out the open strings. 
The ${\mathbb Z}_n^4$-unbroken vacuum can exist only if the twist is strong enough so that the difference of classical potential energies, 
$\beta N^2\left(1-\cos(2\pi/n)\right)\sim \beta N$, is larger than the entropy factor. 
Hence the critical value is given by $\beta_{c,L}\sim N$. 

In the noncommutative Yang-Mills limit, the continuum limit is taken fixing $\theta\sim a^2\sqrt{N}$. 
Although the scaling of the coupling constant in noncommutative Yang-Mills theory is not completely understood because of 
the UV/IR mixing problem, it is reasonable to assume that 
it should scale as in the usual Yang-Mills theory, $\beta(a)\sim -\log a$, because otherwise the agreement in the planar sector is gone. 
Then, it is impossible to take the continuum limit, because $\beta(a)\sim -\log a\sim \log N$ is smaller than $\beta_{c,L}\sim N$ at sufficiently large $N$. 
Essentially the same arguments hold for any bosonic matrix models and noncommutative backgrounds \cite{Azeyanagi:2008bk}. 
This instability is related to an infrared singularity which appears 
in the perturbative calculation of noncommutative Yang-Mills \cite{VanRaamsdonk:2001jd} 
(see also \cite{Armoni:2001uw} for a related argument), 
and hence it is unlikely that it can be avoided by using other nonperturbative formulations.\footnote{
The formulation in \cite{Ambjorn:1999ts} has the same problem. 
}
Note also that many other theories, including the noncommutative generalization of the standard model, fail for the same reason. 

So it seems that I arrived at a conclusion which would upset the participants of this workshop, 
``Workshop on Noncommutative Field Theory and Gravity'' -- 
{\it noncommutative Yang-Mills theory does not exist!} 
At best one can only claim that it exists as an unstable theory, like bosonic string theory. 
In fact, however, the (in-)stability of the background is sensitive to the matter content. 
In the following sections I argue that the instability can be cured in certain cases. 

\section{A cure by supersymmetry}\label{sec:SYM}
In terms of string theory, the instability described above indicates that 
the flux is not strong enough to overcome the attractive force coming from strings between D-branes. 
There I considered only the bosonic degrees of freedom; what happens if fermions are introduced? 
Intuitively, if a few adjoint fermions are added so that the theory becomes supersymmetric, 
then the forces coming from bosonic and fermionic degrees of freedom cancel with each other.\footnote{
Near zero temperature, at nonperturbative level, the attraction can still win \cite{Aoki:1998vn,Hanada:2009hq}. 
However the remaining attractive force is negligible compared to the twist.   
} 

Actual situation can be slightly more complicated, because the flux term and the noncommutative background 
can break supersymmetry.\footnote{
Formally, at $N=\infty$, the Moyal plane can be realized in the supersymmetric matrix models such as the BFSS matrix model 
\cite{Banks:1996vh,deWit:1988wri} 
(the dimensional reduction of $(9+1)$-d ${\cal N}=1$ super Yang-Mills to $(0+1)$-dimension) 
and the IKKT matrix model \cite{Ishibashi:1996xs} 
(the dimensional reduction of $(9+1)$-d ${\cal N}=1$ super Yang-Mills to $(0+0)$-dimension). 
However it is necessary to consider finite-$N$ because we are discussing 
the nonperturbative formulation, and then the flux term is needed.  
} 
There are several `twisted' matrix models which do not break supersymmetry: 
\begin{itemize}

\item
The plane-wave matrix model (or the BMN matrix model) \cite{Berenstein:2002jq}. 

This is a supersymmetric deformation of the BFSS matrix model \cite{Banks:1996vh,deWit:1988wri} ,  
 \begin{eqnarray}
S_{BFSS}=S_b+S_f, 
\end{eqnarray}
where 
\begin{eqnarray}
S_b
=
N\int_0^\beta dt\ {\rm Tr}\left\{
\frac{1}{2}(D_t X_M)^2 
-
\frac{1}{4}[X_M,X_N]^2 
\right\}, 
\end{eqnarray}
and 
\begin{eqnarray}
S_f
=
N\int_0^\beta dt\ {\rm Tr}\left\{
i\bar{\psi}\gamma^{10}D_t\psi
-
\bar{\psi}\gamma^M[X_M,\psi] 
\right\}.   
\end{eqnarray}
The plane-wave deformation is given by \cite{Berenstein:2002jq} 
\begin{eqnarray}
S_{BMN}=S_{BFSS}+\Delta S, 
\end{eqnarray}
where 
\begin{eqnarray}
\Delta S_b
=
N\int_0^\beta dt\ {\rm Tr}\left\{
\frac{\mu^2}{2}\sum_{i=1}^3X_i^2 
+
\frac{\mu^2}{8}\sum_{a=4}^9X_a^2 
+
i\sum_{i,j,k=1}^3\mu\epsilon^{ijk}X_iX_jX_k
\right\}
\end{eqnarray}
and 
\begin{eqnarray}
\Delta S_f
=
\frac{3i\mu}{4}\cdot N\int_0^\beta dt\ {\rm Tr}\left(
\bar{\psi}\gamma^{123}\psi
\right). 
\end{eqnarray}
Here $\epsilon_{ijk}$ is the structure constant of $SU(2)$: $\epsilon_{123}=+1, \epsilon_{213}=-1$ etc. 
The deformation term $\Delta S$ is essentially the twist. 
The fuzzy sphere embedded in $X_{1,2,3}$ is a BPS solution of the equation of motion of this theory. 
About the $k$-coincident fuzzy sphere, the $(1+2)$-dimensional maximally supersymmetric noncommutative Yang-Mills theory can be realized 
\cite{Maldacena:2002rb}. 

As pointed out in \cite{Kim:2003rza}, the plane wave matrix model can be obtained by dimensionally reducing   
4d ${\cal N}$=4 super Yang-Mills on $S^3$ to $(0+1)$-dimension. In the same manner, by dimensionally reducing 
various 4d theories on $S^3$, such as 4d ${\cal N}$=2 and 4d ${\cal N}$=1, the plane wave deformation of 
various supersymmetric matrix models are obtained, by keeping supersymmetry \cite{Kim:2006wg}. 
From them, various 3d noncommutative theories can be obtained.

\item
The flux deformation can also be introduced to the two-dimensional super Yang-Mills\cite{Bonelli:2002mb,Sugiyama:2002tf,Das:2003yq,Hanada:2010gs,Hanada:2010kt,Hanada:2011qx}.  
Fuzzy spheres give stable vacua, and hence noncommutative super Yang-Mills with two spatial noncommutative dimensions can be constructed by taking appropriate limit\cite{Hanada:2010gs,Hanada:2010kt,Hanada:2011qx}.

\item
It is also possible to construct two- and four-dimensional theories starting with zero-dimensional supersymmetric matrix models 
with `twist'; see e.g. \cite{Unsal:2004cf,Ydri:2007ua} for attempts along this direction. 

\end{itemize}

When the regularization breaks the SUSY, the stability of the background is very subtle. 
However, as we will see in Sec.~\ref{sec:multi-adj}, 
it can also be realized by using SUSY-breaking regularizations 
with parameter fine tunings. The ultraviolet and infrared behaviors might be modified then.

\section{A possible cure without supersymmetry}\label{sec:multi-adj}
With supersymmetry, the forces coming from bosonic and fermionic degrees of freedom cancel with each other. 
It means the massless adjoint fermions give repulsive force between eigenvalues (D-branes). 
When more massless adjoint fermions are added, the repulsive force wins and 
eigenvalues repel with each other. At first glance it is a pathological situation: 
eigenvalues spread out indefinitely and hence there is no stable vacuum. 
However it is not a problem if the space is compact, then the uniform distribution 
of eigenvalues can be realized\footnote{
The center symmetry breaking is essentially the deconfinement transition, 
which is related to the Hagedorn growth of the hadronic states. 
It has been argued that the Hagedorn growth is suppressed when adjoint fermions are added \cite{Basar:2013sza}. 
}. 
In terms of matrix model, if the bosonic variables are Hermitian matrices then 
the eigenvalues spread indefinitely and hence the model does not make sense. 
On the other hand, with the unitary variables, eigenvalues are restricted on the unit circle 
and hence there is a natural `infrared cutoff.' Hence the Eguchi-Kawai equivalence holds in this case \cite{Kovtun:2007py}. 
By adding a twist, it should be possible to realize the noncommutative Yang-Mills theory with massless adjoint fermions. 

While the theory with many massless adjoint fermions ($N_f\ge 1$) itself does not seem to be relevant for physics at first glance, 
it is related to $2N_f$-flavor QCD in the following sense. 
Firstly, it is equivalent \cite{Armoni:2003gp,Kovtun:2004bz} to the theories with $2N_f$ fermions in the anti-symmetric representation 
at large-$N$, in the sense that certain correlation functions coincide.  
Secondly, the latter theory can be regarded \cite{Corrigan:1979xf} 
as a large-$N$ limit of QCD, because at $N=3$ the fundamental and the anti-symmetric representations are the same thing. 
A priori there is no reason to favor the fundamental representation rather than the anti-symmetric representation.
Large-$N$ limit with the anti-symmetric fermions is similar to the Veneziano large-$N$ limit ($N_f$ fundamental fermions, 
$N,N_f\to\infty$ with $N_f/N$ fixed), in the sense that the loops of fermions are not suppressed. 

Somewhat surprisingly, the instability can be avoided even when the mass of the adjoint fermion
is as large as the cutoff scale. This was found numerically \cite{Bringoltz:2009kb}, along the way of going to the massless limit.  
(See also \cite{Bringoltz:2011by,Catterall:2010gx,Hietanen:2012ma} for related studies.) 
Soon a simple theoretical explanation was given in \cite{Azeyanagi:2010ne}. 
In a small volume theory, the dynamics of eigenvalues at short distance are approximated by the dimensionally reduced theory,
\begin{eqnarray}
S_{0d}=\frac{1}{g_{0d}^2}{\it Tr}
\left(
-[A_\mu,A_\nu]^2
-
\bar{\psi}_f\left(\gamma^\mu[A_\mu,\psi_f]+m\psi\right)
\right), 
\end{eqnarray} 
where the matrices $A_\mu$ and $\psi_f$ are the zero-modes of the four-dimensional gauge field and fermion and  
the coupling is given by $g_{\rm 0d}^2=g_{\rm 4d}^2/V$, where the volume $V$ is related to the lattice spacing $a$ by $V=a^4$ 
in the case of the single-site lattice. 
The 't Hooft coupling $\lambda_{\rm 0d}=g_{\rm 0d}^2N$, which has the dimension of $({\rm mass})^{4}$, sets the scale of the theory; 
in particular, the fluctuation of the gauge field is of order $\lambda_{\rm 0d}^{1/4}\sim\lambda_{\rm 4d}^{1/4}/a$. 
Compared to this huge fluctuation, the mass term is simply negligible, as long as $m\ll \lambda_{\rm 4d}^{1/4}/a\sim\frac{1}{a\cdot (-\log a)^{1/4}}$. 
Note that the heavy adjoint fermions can be seen near the cutoff scale, and hence the ultraviolet structure is modified. 
Then, due to the UV/IR-mixing, the behavior at deep IR is also modified.  

Although I considered the Eguchi-Kawai model (zero-dimensional theory) and four-dimensional fuzzy space 
with four noncommutative dimensions, the same argument applies to other cases, including 
three-dimensional theory with two noncommutative dimensions (from matrix quantum mechanics) 
and four-dimensional theory with two noncommutative dimensions (from two-dimensional Yang-Mills theory). 

With the heavy adjoint fermions, it is possible to construct various complicated theories, 
for example the noncommutative QCD with U$(N_c)$ gauge group and $N_f$ flavors; 
we only have to construct the U$(N_c)\times U(N_f)$ theory with bifundamental fermions, and then un-gauge U$(N_f)$ \cite{Hanada:2009hd}.  
In the same way, it is possible to realize quiver gauge theories on noncommutative space. 

\section{Conclusion and outlook}\label{sec:outlook}
I have discussed whether the matrix model formulation of the noncommutative Yang-Mills theory can work. 
In general, it depends on the field content. I have shown that very heavy adjoint fermions near the cutoff scale
can make the formulation work, by stabilizing the background matrix configurations. 
As pointed out in \cite{VanRaamsdonk:2001jd}, the instability of the noncommutative background in the bosonic non-commutative Yang-Mills 
corresponds to the tachyonic instability in the perturbation theory arising from the UV/IR mixing. 
The massive adjoint fermions remove those singularities. 
This procedure modifies the theory near UV and IR cutoff scale. 

As physics itself, it would not be a big change, 
because the energy scale of interest is the noncommutativity scale which is far separated from UV and IR cutoff scales. 
For phenomenological model buildings, one can consider any kind of matter content, 
but it is crucial to introduce the right UV completions.  
From more formal theoretical point of view, it is important to study theories with adjoint fermions. 
The stability of the background suggests that the IR singularity is rather harmless and the theory seems to be well-defined.  
Hence these theories with adjoint fermions would be the best ones for theoretical considerations, for example on the renormalizability.

\section*{Acknowledgement}
I would like to thank Tatsuo Azeyanagi, Hidehiko Shimada and Mithat Unsal for discussions and comments.

\end{document}